\documentstyle[aps,epsf,amsmath,amsfonts,bm]{revtex}
\def\be{\begin{eqnarray}}
\def\ee{\end{eqnarray}}
\def\beq{\begin{equation}}
\def\eeq{\end{equation}}

\def\p{\partial}

\def\({\left (}
\def\){\right )}
\def\[{\left [}
\def\[{\right ]}

\bmdefine{\bmj}{ {\bm{j}} }
\bmdefine{\bmB}{ {\bm{B}} }
\bmdefine{\bmF}{ {\bm{F}} }
\bmdefine{\bmL}{ {\bm{L}} }
\newcommand{\bbA}{ {\mathbb{A}} }
\newcommand{\hh}{ {\Hat{h}} }

\newcommand{\hA}{ {\Hat{A}} }

\newcommand{\hR}{ {\Hat{R}} }

\newcommand{\heps}{ {\Hat{\epsilon}} }

\newcommand{\calA}{ {\mathcal{A}} }
\newcommand{\calD}{ {\mathcal{D}} }
\newcommand{\calF}{ {\mathcal{F}} }

\newcommand{\calH}{ {\mathcal{H}} }
\newcommand{\calN}{ {\mathcal{N}} }

\newcommand{\calS}{ {\mathcal{S}} }

\newcommand{\doty}{ {\Dot{y}} }
\newcommand{\dotz}{ {\Dot{z}} }
\newcommand{\dotX}{ {\Dot{X}} }

\newcommand{\tilf}{ {\Tilde{f}} }
\newcommand{\tilg}{ {\Tilde{g}} }
\newcommand{\tilh}{ {\Tilde{h}} }
\newcommand{\tilr}{ {\Tilde{r}} }

\newcommand{\tilG}{ {\Tilde{G}} }

\newcommand{\tilOmega}{ {\Tilde{\Omega}} }
\bmdefine{\bmnabla}{ {\bm{\nabla}} }
\newcommand{\tilhinv}{ {(\tilh^{-1})} }
\newcommand{\tilginv}{ {(\tilg^{-1})} }
\begin{document}
%
\begin{titlepage}
\bigskip
\rightline{}
\rightline{hep-th/0502118}
\bigskip\bigskip\bigskip\bigskip
\centerline{\Large\bf
{Mirror effect induced by the dilaton field on the Hawking radiation}}
\bigskip\bigskip
\bigskip\bigskip

\centerline{\large
Kengo Maeda\footnote{e-mail:kmaeda@kobe-kosen.ac.jp} and
Takashi Okamura\footnote{e-mail:okamura@ksc.kwansei.ac.jp}
}
\bigskip\bigskip
\centerline{\em ${}^*$
Department of General Education, Kobe City College of Technology }
\centerline{\em
8-3 Gakuen-higashi-machi, Nishi-ku, Kobe 651-2194, Japan}
\bigskip
\centerline{\em
${}^\dagger$ Department of Physics, Kwansei Gakuin University,
Sanda 669-1337, Japan}
\bigskip\bigskip
\begin{abstract}
We discuss the string creation
in the near-extremal NS1 black string solution.
The string creation is described by an effective field equation
derived from a fundamental string action coupled to the dilaton field
in a conformally invariant manner.
In the non-critical string model the dilaton field causes
a timelike mirror surface outside the horizon
when the size of the black string is comparable to the Planck scale.
Since the fundamental strings are reflected by the mirror surface,
the negative energy flux does not propagate across the surface.
This means that the evaporation stops just before
the naked singularity of the extremal black string appears
even though the surface gravity is non-zero in the extremal limit.
\end{abstract}

\end{titlepage}

\baselineskip=18pt

\setcounter{equation}{0}
\section{Introduction}
It is generically agreed that quantum gravity might remove
pathological objects in classical general relativity like singularity.
One of the difficulties in quantizing gravity is
the non-renormalizability of itself.
So, we expect that string theory is the best of candidates
for consistent quantum gravity theory
since string theory is free of the ultraviolet divergences.
It was shown that the first-quantized string is well defined on
orbifold and weak curvature singularities~\cite{dixon,horowitzsteif}.
However, every curvature singularity is not resolved
at the first-quantization level~(See ref.~\cite{horowitzsteif})
since mass of the first-quantized string gets excited infinitely
for strong curvature singularities.
Such a problem could be resolved by
constructing string field theory describing pair creation of strings.

Lawrence and Martinec~\cite{lawrence} derived
an effective string field equation from the fundamental string action
in which the coupling term $\alpha'R\Phi$
between curvature of the world-sheet $R$ and the dilaton field $\Phi$
is absent.
Based on the equation the pair creation rate was calculated
on various cosmological models~\cite{lawrence,gubser1,gubser2}.
When the dilaton field is almost constant,
this coupling term is negligibly small as this is $O(\alpha')$.
The coupling term, however, could be dominant near singularity
where the dilaton field diverges.
In this paper we investigate the effect of the dilaton field
on the string creation in near-extremal black strings.
The effective field equation describing the string creation is derived
from a string action conformally coupled to the dilaton field.
To simplify the equation,
we dimensionally reduce the fundamental string to a particle
by omitting the string excitations and by compactification.

In Einstein Gravity charged black holes are described by
Reissner-Nordstr\"om solution, where the temperature approaches zero
in the extremal limit.
So, the extremal black hole in Einstein Gravity
is stable against the Hawking radiation.
In string theory there are black branes which have
finite temperature and a singular horizon in the extremal limit.
This leads physicists to imagine that
the singular horizon, or naked singularity appears by evaporation.
This expectation was partially verified in~\cite{kogamaeda}
by investigating dilatonic black holes.

We investigate the evaporation process of
five-dimensional NS1 black string
with Kaluza-Klein charge~(W-charge) by the string creation.
When the W-charge is zero, the temperature of the black string
becomes infinite in the extremal limit
and hence the quasi-adiabatic approximation breaks down.
When the W-charge is non-zero,
the temperature is non-zero but finite in the limit.
So, we can test whether the naked singularity appears
within the quasi-adiabatic approximation.

For the non-critical string model,
we find that the effective field equation becomes singular
at a timelike surface outside the horizon
when the black string is comparable to the string scale
$\sqrt{\alpha'}$.
We show that the effective field equation has a unique solution
for the initial data with finite energy so that
the natural function space of initial data is the Sobolev space $H^1$.
The well-posedness of massive scalar field was shown
in some interesting spacetimes for the field
in the Sobolev space~\cite{IH}.
Since the string current passing through the timelike surface is zero
for each mode in the Sobolev space,
this surface plays a role of a mirror surface.
So, the negative energy flux does not propagate across the surface,
implying that the evaporation by the string creation stops
before the occurrence of the naked singularity.
For the critical string model,
we find that the effective field equation does not become singular
until the naked singularity appears,
implying that the black string continues to evaporate.
In Sec.\ref{sec:conclusion},
we will argue that the occurrence of the naked singularity could
be prevented if we consider the time dependence of the dilaton field.

For completeness, we also investigate the stability of
near-extremal NS1 black string in which
temperature is zero in the extremal limit.
In general it is possible for the zero temperature black hole
to evaporate by the spontaneous discharge process,
as shown in~\cite{ref:Gibbons}.
We show that the evaporation continues
unless the W-charge vanishes.
Since the mass-loss rate is smaller than the charge-loss rate,
it leaves the extremal state.
So, the naked singularity does not occur.

This paper is organized as follows.
In the next section,
we formulate string creation based on the particle picture.
In section \ref{sec:NS1-W},
we discuss the evaporation process and superradiance
by the string creation.
Section \ref{sec:conclusion} is devoted to conclusion and discussions.
\section{Formulation of string creation based on particle picture}
We study the string creation problem in the particle picture,
which is obtained by dimensionally reducing the string action
with the dilaton coupling term $\alpha'\, R\, \Phi$
to the particle one.
If we dimensionally reduce the action with the coupling term
$\alpha'\, R\, \Phi$ directly,
then we are faced with a bit technical difficulty.
It comes from the breaking of the Weyl invariance
by the coupling term $\alpha'\, R\, \Phi$ at the classical level.
Indeed, in the conformal gauge $h_{ab} = e^{2 \chi}\, \eta_{ab}$,
the coupling term produces
\begin{align}
  & \alpha' \int d^2\sigma~\sqrt{-h}\, R\, \Phi
  = 2\,\alpha'\, \int d^2\sigma~(\partial \Phi) \cdot (\partial \chi)~.
\label{eq:trouble}
\end{align}
The interaction term shows that the field $\chi$
to be unphysical gauge mode becomes dynamical degrees of freedom.
Of course, the reason is that
we do not incorporate quantum corrections.
In order to overcome this difficulty
by incorporating the quantum corrections,
we translate the Polyakov-type action with $\alpha'\, R\, \Phi$ term
into the action with Weyl invariant dilaton coupling term
which gives the same partition function as
the original Polyakov-type action.
In the appendix \ref{sec:model}, we explain it in detail.

The Lorentzian action of a test string propagating
in $D$-dimensional target space,
which corresponds to the Euclidean action (\ref{eq:def-calS})
is given by
\begin{align} 
  & \calS
  = \frac{-1}{4 \pi \alpha'} \int d^2\sigma~\sqrt{-h\,}~
     \left( h^{ab}\, \tilG_{MN}(X) + \epsilon^{ab}\, B_{MN}(X) \right)
       \partial_a X^M \partial_b X^N~,
\label{eq:full-action} \\
  & \tilG_{MN}(X)
  := G_{MN}(X) + \gamma\, \alpha'\,
            \partial_M \Phi(X) \cdot \partial_N \Phi(X)~,
\label{eq:def-tilG}
\end{align}
where $\gamma = 6/(D - D_{\text{crit}}+1)$
and we drop \lq\lq hat\rq\rq\, from the fiducial metric
and use the capital Latin letters as the indices of the target space.
\subsection{Kaluza-Klein reduction of the test string}
Hereafter, we would like to consider a test string
in $(d+1)$-dimensional spacetimes
obtained by toroidal compactification.
And then, we assume that the $(d+1)$-dimensional spacetimes
have one Killing direction and
a test string extends along the direction only.

Concretely, we consider the background fields as follows;
\begin{align} 
  & ds^2 = G_{MN}~dX^M dX^N
  = ds_{d+1}^2 + \sum_{M=d+1}^{D-1} (dX^M)^2~,
\label{eq:targetD} \\
  & ds_{d+1}^2
  = g_{\mu\nu}(x) dx^\mu dx^\nu
  + k^2(x) \left( dX^d + \hA_\mu(x)\, dx^\mu \right)^2~,
\label{eq:bg_metric} \\
  & \bmB = \calA_\mu(x)\, dx^\mu \wedge dX^d~,
& & \Phi = \Phi(x)~,
\label{eq:bg_dilaton}
\end{align}
where $\mu, \nu = 0, 1, \cdots, d-1$, and then the \lq\lq effective
metric\rq\rq\, $\tilG_{MN}$ becomes
\begin{align} 
  & \tilG_{\mu\nu} = \tilg_{\mu\nu}(x)
  := g_{\mu\nu}(x) + \gamma\, \alpha'\,
            \partial_\mu \Phi(x) \cdot \partial_\nu \Phi(x)~,
& & \tilG_{dd} = G_{dd} = k^2(x)~,
\label{eq:bg_tilG}
\end{align}
and so on.
In the target spacetime with its metric (\ref{eq:bg_metric}),
we would like to investigate the dynamics of the the test string
which extends along the $X^d$-axis,
\begin{align}
  & x^\mu = z^\mu(\tau)~, & & X^d = - m\, \sigma + y(\tau)~,
& & X^M = \text{fixed} \hspace{0.5cm} (M=d+1, \cdots, D-1)~.
\label{eq:ansatz-string}
\end{align}
From the canonical form
$P_M\, \dotX^M = p_\mu\, \dotz^\mu + P_d\, \doty$,
we know that $P_d$ is the canonically conjugate momentum of $y$.
Furthermore, the variable $y$ becomes cyclic coordinate,
since $X^d$ is the Killing direction.

In the ansatz, we can take the gauge conditions
for the world sheet reparametrization,
\begin{align}
  & h_{00} = - \calN^2(\tau)\, e^{2 \chi(\tau) }~,
& & h_{11} = e^{2 \chi(\tau) }~,
& & h_{01} = 0~,
\label{eq:ansatz-WSmetric}
\end{align}
and we can obtain the effective action
for the Kaluza-Klein (KK) compactified test string,
\begin{align}
  & \calS = \kappa \int d\tau \left(
    \frac{1}{2\, \calN}\, \tilg_{\mu\nu}\, \dotz^\mu\, \dotz^\nu
  + \frac{k^2}{2\, \calN}\, \left( \doty + \hA_\mu \dotz^\mu \right)^2
  + m\, \calA_\mu \dotz^\mu - \calN\, \frac{m^2\, k^2}{2} \right)~,
\label{eq:effective_action}
\end{align}
where $\kappa := l/2 \pi \alpha'$ and we take $0 \le \sigma \le l$.

Since the variable $y$ is cyclic coordinate,
the canonically conjugate momentum of $y$ is a constant
$p_y = (\kappa\, k^2/\calN) ( \doty + \hA_\mu \dotz^\mu ) =: q$,
and then we can simplify the action (\ref{eq:effective_action})
with the Rauth function,
\begin{align}
   \calS'
  &= \int d\tau \Bigg[~\frac{\kappa}{2\, \calN}\, 
      \tilg_{\mu\nu}\, \dotz^\mu\, \dotz^\nu
  + \bbA_\mu(z) \dotz^\mu - \frac{\calN}{2 \kappa}\, V(z)~\Bigg]~,
\label{eq:Rauthian}
\end{align}
where
\begin{align}
  & \bbA_\mu(x)
  := (\kappa\, m)\, \calA_\mu(x) + q\, \hA_\mu(x)~,
& & V(x)
  := (\kappa\, m)^2\, k^2(x) + \left( \frac{q}{k(x)} \right)^2~.
\label{eq:def-V}
\end{align}
From the former equation in Eq.(\ref{eq:def-V}),
we know that this KK-compactified test string
carries the Kalb-Ramond charge $\kappa m$ and the W-charge $q$.
\subsection{Quantization}
Hereafter, we denote the string position coordinates by $x$.
In order to quantize the system,
we should make Hamiltonian formulation.
Following the standard procedure, we have a Hamiltonian constraint
\begin{align}
   \calH'
  &:= \frac{1}{2 \kappa}\, \left[~\big(\, p_\mu - \bbA_\mu\, \big)\,
     \tilginv^{\mu\nu}\, \big(\, p_\nu - \bbA_\nu\, \big)
  + V(x)~\right] \approx 0~,
\label{eq:def-constraint}
\end{align}
where $\tilginv^{\mu\nu}$ is the inverse matrix of $\tilg_{\mu\nu}$.

According to the standard procedure of the canonical quantization,
we take a wave function $\Psi(x)$ and
replace the momenta with the derivatives, and then,
we obtain the equation of motion,
\begin{align}
  & 0 = 2 \kappa\, \calH'\, \Psi(x)
  = - \left[~\calD_\mu\, \tilginv^{\mu\nu}\, \calD_\nu
           - V(x)~\right]\, \Psi(x)~,
& & \calD_\mu := \nabla_\mu - i\, \bbA_\mu~,
\label{eq:KG_eqn}
\end{align}
where we adopt the covariant derivative with respect to
the original metric $g_{\mu\nu}$ as momentum operators,
$p_\mu \rightarrow -i\, \nabla_\mu$ and adopt the operator ordering
by which there exists the conserved inner product.
Indeed, for the current $J^\mu$ defined by
\begin{align}
  & J^\mu( \Psi_1,\, \Psi_2 )
  := i\, \tilginv^{\mu\nu} \left[~\Psi_1^*~\calD_\nu \Psi_2
  - \left( \calD_\nu \Psi_1 \right)^*~\Psi_2~\right]~,
\label{eq:def-tilJ}
\end{align}
the equation $\nabla_\mu J^\mu = 0$ holds
and we can define the conserved inner product as
\begin{align}
   \big( \Psi_1,\, \Psi_2 \big)
  &:= \int_{\Sigma} d\Sigma~n_\mu\, J^\mu( \Psi_1,\, \Psi_2 )
  = i \int_{\Sigma} d\Sigma~n_\mu\, \tilginv^{\mu\nu} \left[~
        \Psi_1^*~\calD_\nu \Psi_2
    - \left( \calD_\nu \Psi_1 \right)^*~\Psi_2~\right]~.
\label{eq:def-KG_IP}
\end{align}

The equation (\ref{eq:KG_eqn}) is also derived from the action
\begin{align}
   S_\Psi
  &= - \int d^{d}x~\sqrt{-g}~\left[~\tilginv^{\mu\nu}\,
     \big( \calD_\mu\, \Psi \big)^*\, \big( \calD_\nu \Psi \big)
  + V(x)~\vert\, \Psi\, \vert^2~\right]~.
\label{eq:action-Psi}
\end{align}
By regarding the field $\Psi(x)$ as the quantized field operator and
using the inner product (\ref{eq:def-KG_IP}),
we could formally make the second quantization
and argue the \lq\lq string creation\rq\rq\,
in the same manner as the particle creation.

Hereafter, we are interested in static cases as
\begin{align} 
  & g_{\mu\nu}\, dx^\mu dx^\nu
  = - N^2 dt^2 + h_{ij}~dx^i\, dx^j~,
& & \bbA_\mu\, dx^\mu = \bbA_t\, dt~,
& & \nabla_\mu \Phi = D_\mu \Phi~,
\label{eq:static-bg}
\end{align}
where $D_\mu$ is the covariant derivative
with respect to the spatial metric $h_{\mu\nu}$.
In this case, the field equation (\ref{eq:KG_eqn}) becomes
\begin{align} 
  & \calD_t^2 \Psi = - K\, \Psi~,
& & K
  := - N\, \left[~D_i \big( N\, \tilhinv^{ij} D_j \big)
  - N\, V~\right]~,
\label{eq:def-hK}
\end{align}
and then, 
the inner product (\ref{eq:def-KG_IP}) suggests
a natural inner product for functions on the spatial hypersurface
$\Sigma$ as
\begin{align} 
  & \left\langle u,\, v \right\rangle
  := \int_\Sigma \frac{d\Sigma}{N}~u^*~v~.
\label{eq:def-spatial_IP}
\end{align}

For later convenience, we introduce another inner product
on the spatial surface, the Sobolev one
\begin{align} 
   \langle u,\, v \rangle_{\text{Sob}}
  &:= q^2\, \int d\Sigma~\frac{ u^*\, v }{N}
  + \int d\Sigma~N\, \left[~\tilhinv^{ij}\, (D_i u)^*\, (D_j v)
      + V\, u^*\, v~\right]~,
\label{eq:def-Sob_IP}
\end{align}
where $q$ is some real constant.
We note that the operator $K$ is hermitian
with respect to not only the inner product (\ref{eq:def-spatial_IP}),
but also the Sobolev inner product,
\begin{align}
  & \langle u,\, K v \rangle_{\text{Sob}}
  - \langle K u,\, v \rangle_{\text{Sob}}
\nonumber \\
  &= \oint_{\partial \Sigma} dS_i~N\, \tilhinv^{ij}\,
    \left[~(D_j u)^*\, \left\{~(q^2 + K )\, v~\right\}
     - \left\{~(q^2 + K )\, u~\right\}^*\, (D_j v)~\right] = 0~.
\label{eq:hermiticity-hK-Sob}
\end{align}
%
\section{The evaporation process on NS1-W solution}\label{sec:NS1-W}
The metric of the target spacetime and the dilation field
for the NS1-W solution
\cite{ref:NS1-W}
are given by
\begin{align} 
  & ds_{d+1}^2 = - \frac{f(r)}{ H_\alpha(r) H_\beta(r) }\, dt^2
  + f^{-1}(r)\, dr^2 + r^2 d\Omega_{d-2}
  + \frac{ H_\beta(r) }{ H_\alpha(r) }\,
      \left( dX^d + \hA_\mu(r)\, dx^\mu \right)^2~,
\label{eq:NS1W-metric} \\
  & e^{-2\Phi} = H_\alpha(r)
  = 1 + \left( \frac{r_0}{r} \right)^{d-3}\, \sinh^2\alpha~,
\label{eq:NS1W-dilaton}
\end{align}
where $f(r) = 1 - (r_0/r)^{d-3}$.
The various quantities are summarized in the appendix
\ref{sec:NS1-W-App}.

We denote the charges related to the Kalb-Ramond field $\calA_\mu$ 
and to the KK U(1) field $\hA_\mu$
by $Q_\alpha$ and $Q_\beta$, respectively.
The extremal limit is that the horizon position
$r_0 \rightarrow 0$, $\alpha, \beta \rightarrow \infty$
keeping $Q_\alpha, Q_\beta$ fixed.
The ADM mass $M$ is bounded below $M \ge Q_\alpha + Q_\beta$.

The Hawking temperature and the entropy in the extremal limit
are given by
\begin{align}
  & S_{\text{BH}} \sim r_0~\sqrt{Q_\alpha Q_\beta}~,
& & T_{\text{BH}}
  \sim \frac{ r_0^{d-4} }{ \sqrt{Q_\alpha Q_\beta} }~,
\label{eq:T_BH-ext}
\end{align}
so that, in the extremal limit, the temperature approaches
a finite value $T_H\sim 1/\sqrt{Q_\alpha Q_\beta}$ for $d=4$
and the zero temperature for $d \ge 5$.
Therefore, it is very interesting to consider the evaporation process
of near extremal five dimensional ($d=4$) NS1-W solution.

In the background (\ref{eq:NS1W-metric})-(\ref{eq:NS1W-dilaton}),
the metric $\tilde{g}_{\mu\nu}$ appeared on the first quantization
of the fundamental string is given by
\begin{align}
  & \tilg_{\mu\nu} dx^\mu dx^\nu
  = - \frac{f(r)}{ H_\alpha(r) H_\beta(r) }\, dt^2
  + \tilf^{-1}(r) dr^2 + r^2 d\Omega_{d-2}~,
\label{eq:tilded-KK-metric} \\
  & \tilf(r)
  := f(r)~\left[~1 + \frac{\gamma\, \alpha'}{4}\, f\,
        \left( \frac{H'_\alpha}{H_\alpha} \right)^2~\right]^{-1}~,
\label{eq:def-tilf}
\end{align}
and the KK-compactified test string
couples to $U(1)$ gauge field $\bbA_\mu$ and the potential $V$,
\begin{align}
  & \bbA_\mu dx^\mu
  = - \left( (\kappa\, m)\,
             \frac{\sinh\alpha\, \cosh\alpha}{H_\alpha(r)} 
  + q\, \frac{ \sinh\beta\, \cosh\beta}{H_\beta(r)}
      \right) \left( \frac{r_0}{r} \right)^{d-3}~dt~,
\label{eq:KK_UI} \\
  & V
  = (\kappa\, m)^2\, \frac{H_\beta(r)}{H_\alpha(r)}
  + q^2\, \frac{H_\alpha(r)}{H_\beta(r)}~,
\label{eq:KK_V}
\end{align}
where a dash means the derivative with respect to $r$.

As shown in the previous section,
the only difference between usual complex scalar field equation
and our effective field equation (\ref{eq:KG_eqn}) derived from the first quantization
is the difference between $g$ and $\tilde{g}$
appeared in the kinetic terms.
Since $\sqrt{\alpha'}$ is the string scale,
the corrections from the dilaton field are negligibly small
when the the size of the black hole $r_0$ is much larger than
the string scale.
However, they are important when the horizon becomes
comparable to the string scale.

Anyway, since the background is static and spherically symmetric,
for simplicity, we consider spherically symmetric solution
(S-wave solution)
$\Psi = e^{-i \omega t}~\psi(r)/r^{(d-2)/2}$,
and then, Eq.(\ref{eq:KG_eqn}) reduces
to the radial equation
\begin{align}
  &0 = \left( \frac{d^2}{d\tilr_*^2}
  + \frac{\tilf}{f}\, \left( \omega + \bbA_t \right)^2
  - \tilf~U(r) \right)~\psi~,
\label{eq:KG_eqn-radial} \\
  & U
  := \left( \frac{\kappa\, m}{H_\alpha} \right)^2
  + \left( \frac{q}{H_\beta} \right)^2
  + \frac{d-2}{2\, r}~\frac{1}{ r^{(d-4)/2} }\,
    \frac{1}{ \sqrt{H_\alpha H_\beta} }
      \left( \frac{r^{(d-4)/2}\, \tilf}{ \sqrt{H_\alpha H_\beta} }
      \right)'~,
\label{eq:def-U}
\end{align}
where we introduce the modified tortoise coordinate $\tilr_*$
\begin{align}
  & \tilr_* := \int dr~\sqrt{H_\alpha H_\beta}/\tilf~.
\label{eq:def-tilr_star}
\end{align}
%
\subsection{evaporation of near extremal black string
with finite temperature}\label{sec:non-critical}
We shall consider the evaporation of
the five-dimensional NS1-W solution,
which has finite temperature in the extremal limit.

First we investigate the process by the non-critical string model,
$\gamma<0$
\footnote{
Strictly speaking, we need to use a non-critical black string solution
instead of the NS1-W solution for consistency
with the non-critical string model.
For our best knowledge, the non-critical black string solution
is not yet obtained. So, we simply use the NS1-W solution
to investigate the creation of the non-critical strings.
}.
As the black string approaches the extremal state,
$H_\alpha$ and $\tilde{f}$ can be approximated as
\be
H_\alpha\sim \frac{r_0}{r}\sinh^2\alpha, \qquad
\tilde{f}=f\left(1+\frac{\gamma{\alpha}' f}{4r^2}\right)^{-1}
\ee
near the horizon.
When the size of the black hole becomes
$r_0 = (1/3) \sqrt{ \vert \gamma \vert\, \alpha'/3}$,
the effective field equation (\ref{eq:KG_eqn-radial}) is singular
at $r_1=(1/2) \sqrt{ \vert \gamma \vert \alpha'/3}$,
since $\tilde{f}$ diverges at $r_1$ as $\tilde{f}\sim 1/(r-r_1)^2$.

Let us consider a boundary condition on it.
As a physically reasonable condition,
we shall impose that the total energy $E$ is finite:
\be
\label{finite-E}
E&:=&\int d\Sigma~n_\mu\xi_\nu T^{\mu\nu}
= \int d\Sigma~N\left[~\frac{|\calD_t \Psi|^2}{N^2}
+\tilde{h}^{ij}(D_i\Psi)^*(D_j\Psi)+V|\Psi|^2~\right]<\infty,
\ee
where the energy-momentum tensor is usually defined by
$T^{\mu\nu} := (2/\sqrt{-g})(\delta S_\Psi/\delta g_{\mu\nu})$.
Since $\tilde{h}^{ij}\sim \tilde{f}$, the above condition
implies that for $r \sim r_1$,
\be
\label{bc}
\Psi\sim (r-r_1)^n, \qquad n>3/2.
\ee

Near the singularity, the wave-equation (\ref{eq:KG_eqn-radial})
is approximately written by
\be
\label{one-potential}
0=\left(\frac{d^2}{d{\tilde{r}_*}^2}
+c{\tilde{r}_*}^{-5/3}\right)\psi(r),
\qquad \mbox{for}\qquad  r>r_1
\ee
where $c$ is a positive number and $r_* \sim (r-r_1)^3$.
The two-independent solutions are
expressed by Bessel functions:
\be
\label{mode-sol}
\psi_1&\sim& \sqrt{\tilde{r}_*}J_3(6\sqrt{c}\tilde{r}_*^{1/6})
\sim  \tilde{r}_*\sim (r-r_1)^3, \nonumber \\
\psi_2&\sim&\sqrt{\tilde{r}_*}N_3(6\sqrt{c}\tilde{r}_*^{1/6})
\sim 1+a\tilde{r}_*^{1/3}\sim 1+a(r-r_1),
\ee
where $a$ is a number.
So, $E$ is finite for $\psi_1$,
while $E$ diverges for $\psi_2$.
This implies that the only $\psi_1$ mode is admitted for the singular
boundary. Using Eq.~(II.28), one can easily check that $\psi_1$
produces no string current on the boundary:
\be
\label{current}
n_\mu J^\mu\sim \tilde{f}\psi_1\p_r\psi_1\sim (r-r_1)^3\to 0.
\ee
So, the surface at $r=r_1$ plays a role of a reflective mirror.
This means that the evaporation process ceases when the horizon reaches
the string scale and that no naked singularity appears
for the non-critical string case
\footnote{When the size of the black string becomes
$r_0 < (1/3) \sqrt{|\gamma| \alpha'/3}$, $\tilf$ can be negative
in the region $r \in (r',\, r'')$ outside the horizon,
where $r_0 < r' < r_1 < r''$.
So, we have \lq\lq two time\rq\rq\, $\tilg_{tt},\, \tilg_{rr} < 0$
in the region.
However, by the same argument in the main text,
we can show that the surface at $r=r''$ also plays a role of
a reflective mirror owing to the behavior of $\tilf$ at $r=r''$
as $\tilf \sim 1/(r-r'')$ and it makes the \lq\lq two time\rq\rq\,
region separate from the asymptotic region $r'' < r$.
}.

It is worthwhile to comment another meaning of the condition
(\ref{finite-E}).
Because $\langle \Psi,\, \Psi \rangle_{\text{Sob}}
< \text{(const.)} \times E < \infty$,
the condition (\ref{finite-E}) imposes that
the allowed states belong to the Sobolev space $H^1$
with the inner product (\ref{eq:def-Sob_IP}).
It is easy to check that $\psi_1 \in H^1$ and
$\psi_2 \notin H^1$.
This means that the operator $K$ is essentially self-adjoint
in the Sobolev space, so that the initial value problem of $\Psi$
is well-posed
\cite{ref:HawkingEllis}
\footnote{On the other hand, by the method similar to that noted below,
we can show that the operator $K$ is not essentially self-adjoint
in the linear function space with the square integrability $L^2$,
so that the evolution is not unique.}.
Essential self-adjointness of $K$ is proved by showing that
we have no solution in $H^1$ to the each eigenvalue equation
$K \Psi = \pm\,i\, \Psi$
\cite{ref:ReedSimon},
which is reduced to the similar equation
to Eq.(\ref{eq:KG_eqn-radial}),
\begin{align}
  & 0 = \left( \frac{d^2}{d\tilr_*^2}
  \pm\, i\, \frac{\tilf}{f} - \tilf~U(r) \right)~\psi~.
\label{eq:eigen_eqn}
\end{align}
Due to the singular behavior of the last term of $U$ near $r \sim r_1$,
both equations become Eq.(\ref{one-potential}) and the solutions
behave as Eq.(\ref{mode-sol}), so that the solutions in $H^1$
of Eq.(\ref{eq:eigen_eqn}) should satisfy the boundary condition
$\Psi(r=r_1)=\Psi(r_*=0)=0$ and also $\Psi(r_*=\infty)=0$.
We will have no solution of the eigenvalue equation
(\ref{eq:eigen_eqn}) satisfying the above boundary condition,
because we have no room to adjust the eigenvalues.
Thus, the operator $K$ will be essentially self-adjoint
in the Sobolev space.

Secondly, we consider the evaporation process
by the critical string model, $\gamma>0$.
As easily checked, the effective potential $U$ in Eq.~(\ref{eq:def-U})
is bounded above by the charges $Q_\alpha$ and $Q_\beta$ as
\be
 U \lesssim 1/Q_\alpha Q_\beta
\ee
in the extremal limit, implying that the evaporation continues
until naked singularity appears in the extremal limit.
What has to be noticed is that we implicitly assumed that
the NS1-W black string eventually evaporates and then
the quasi-adiabatic approximation is valid  for all time.
We will argue that this condition is violated
at the final stage of the evaporation and that
a timelike mirror surface would be formed even
in the critical string model in section \ref{sec:conclusion}.
\subsection{evaporation of the near extremal black string solution
with zero temperature}
In this subsection, we discuss whether or not
the black string solution in which temperature is zero
in the extremal limit is stable against the string creations.
Since the string creation caused by gravitational effect
is negligibly small compared with the superradiance or
the spontaneous discharge near the extremal limit,
the stability is determined by the spontaneous discharge process only.

The condition for the spontaneous discharge is that
there exist the modes with negative phase velocity.
From the wave equation~(\ref{eq:KG_eqn-radial}) near the horizon
\begin{align}
  & 0 = \left( \frac{d^2}{d\tilr_*^2} + \Omega^2 _\omega
  + O( f ) \right) \psi_{\omega \lambda}~,
& & \Omega_\omega
  := \omega - \big[~(\kappa m)\, \tanh\alpha + q\, \tanh\beta~\big]~,
\label{eq:def-Omega}
\end{align}
they should satisfy
\begin{equation}
   \omega
  < (\kappa m)\, \tanh\alpha + q\, \tanh\beta =: \omega_{\text{SR}}~.
\label{eq:cond_horizon}
\end{equation}
From the wave equation near infinity
\begin{align}
  & 0 = \left( \frac{d^2}{dr^2} + \tilOmega^2_\omega
   + O( r^{-2} ) \right) \psi_{\omega \lambda}~,
& & \tilOmega_\omega
  := \sqrt{ \omega^2 - \left[~(\kappa\, m)^2 + q^2~\right] }~,
\label{eq:def-tilOmega}
\end{align}
the modes should also satisfy
\begin{equation}
  \omega^2 > (\kappa\, m)^2 + q^2 =: \mu^2
\label{eq:cond_infty}
\end{equation}
for the wave propagation to infinity.
Combining the above two inequalities, we obtain
the stability condition against the spontaneous discharge process,
$\mu > \omega_{\text{SR}}$, which is automatically satisfied
by the BPS-like bound $\mu > (\kappa\, m) + q \ge \omega_{\text{SR}}$
and is equivalent to
\begin{align}
   \frac{ (\kappa\, m)^2 }{\cosh^2\alpha} + \frac{q^2}{\cosh^2\beta}
  > 2 (\kappa m)\, q\, \tanh\alpha\, \tanh\beta~.
\label{eq:cond_superR}
\end{align}
So, provided that this condition (\ref{eq:cond_superR}) holds,
the extremal black string solution with zero temperature
is stable against the string creation.

In the extremal limit $\alpha, \beta \to \infty$,
the condition (\ref{eq:cond_superR}) is automatically broken
unless $q=0$, so that the black string solution with the W-charge
is unstable.
By the standard procedure~\cite{ref:Gibbons},
the mass-loss rate and the charge-loss become
\begin{align}
  & \frac{dM}{dt}
  = - \int^{\omega_{\text{SR}}}_\mu d\omega~\omega
      \left( \left\vert\, R_\omega\, \right\vert^2 - 1 \right)~,
\label{eq:dM-dt} \\
  & \frac{dQ_\alpha}{dt}
  = - (\kappa m) \int^{\omega_{\text{SR}}}_\mu d\omega~
      \left( \left\vert\, R_\omega\, \right\vert^2 - 1 \right)~,
& & \frac{dQ_\beta}{dt}
  = - q \int^{\omega_{\text{SR}}}_\mu d\omega~
      \left( \left\vert\, R_\omega\, \right\vert^2 - 1 \right)~,
\label{eq:dQ_beta-dt}
\end{align}
where $R_\omega$ is the reflection amplitude
and for the superradiance modes (\ref{eq:cond_horizon}),
$\left\vert\, R_\omega\, \right\vert^2 > 1$.

Since $\kappa m + q \ge \omega_{\text{SR}}$ and
$\left\vert\, R_\omega\, \right\vert^2 > 1$
for $\omega_{\text{SR}} > \omega$, we conclude that
\begin{align}
  & 0 > \frac{dM}{dt} > \frac{dQ_\alpha}{dt} + \frac{dQ_\beta}{dt}
& & \Longrightarrow
& & \left\vert\, \frac{dM}{dt}\, \right\vert
  < \left\vert\, \frac{d}{dt}\, ( Q_\beta + Q_\beta)\, \right\vert~.
\label{eq:dischargeII}
\end{align}
This means that the near-extremal black string has more discharge rate
than mass-loss one.
Therefore provided that the quasi-adiabatic approximation is valid,
the over-charge process does not occur
even for the black string solution with the W-charge,
so that the naked singularity at the extremal state does not appear
for the black string solution with zero temperature
in both the critical and non-critical string models.
\section{Conclusion and Discussions}\label{sec:conclusion}
We investigated the effect of dilaton field on string creation
in near-extremal NS1-W black string solutions. 
For a simplicity we assumed that the string extends
along one Killing direction only.
So, it is reduced to a charged particle whose motion
is derived from the effective action~(\ref{eq:effective_action})
in the lower dimensional spacetime. 

As shown in Sec.\ref{sec:non-critical}
the determinant of the \lq\lq effective metric\rq\rq\,
$\tilde{g}_{\mu\nu}$ becomes zero at a timelike surface
outside the horizon when the horizon size is close to the Planck scale
in the non-critical string model.
It follows that the field equation describing the particle creation
becomes singular at the surface.
Assuming that all the states belong to the Sobolev space $H^1$,
we showed that the evolution is uniquely defined for all time and that
there is no particle flux propagating across the surface.
In other words, the black string at the Planck scale
is completely enclosed by the timelike \lq\lq mirror\rq\rq\, surface
and hence the evaporation by the Hawking radiation stops.
Of course, this conclusion is based on perturbative treatment
of strings and the quasi-adiabatic approximation.
Although we need any non-perturbative treatment of string theories
for a complete understanding of the whole evaporation process
due to the formation of the mirror surface at the Planck regime,
it is beyond our scope.

It is worth noting that
we used a five-dimensional NS1 black string solution
as a background spacetime instead of a non-critical solution simply
because no analytical black hole solution
is found in the non-critical string theory.
We expect, however, that the determinant of the \lq\lq effective 
metric\rq\rq\,$\tilde{g}_{\mu\nu}$ becomes zero near curvature 
singularity even in the non-critical solution unless the dilaton field 
is finite there.

Finally we comment on the case for the critical string model
in a five-dimensional NS1 black string solution
under the quasi-adiabatic approximation.
Since the timelike mirror surface is not produced in this case,
the evaporation by the Hawking radiation seems continue
until the naked singularity appears.
Let us consider if the surface is produced or not
when we take into account the time-dependence of the dilaton field.
As easily seen in Eq.~(\ref{eq:NS1W-dilaton}),
the dilaton field diverges at the extremal state.
So, the time derivative would also diverge at the extremal state
if the evaporation is completed within a finite time. 
Since the time component of the metric $\tilde{g}$ changes as
\be
\tilde{g}_{tt}=g_{tt}+\gamma (\p_t\Phi)^2~,
\ee
it would become zero at the last stage 
(note that $\gamma>0$ in the critical string model).
This implies that the a mirror surface
preventing the formation of the naked singularity
is produced provided that the time dependence is taken into account.
This will be discussed in more detail in~\cite{MO}.
\begin{acknowledgments}
It is a pleasure to thank M. Natsuume and S. Yahikozawa
for continuous discussions.
\end{acknowledgments}
\appendix
\section{Derivation of the model action (II.2)}\label{sec:model}
We consider the bosonic string in the background fields,
$G_{\mu\nu}(X)$, $B_{\mu\nu}(X)$ and $\Phi(X)$,
whose action is given by
\footnote{We follow the sign convention of the Polchinski's textbook
\cite{ref:Polchinski_T}.}
\begin{align} 
  & S[X,\, h\,;~G,\, B,\, \Phi]
  = \frac{1}{4 \pi \alpha'} \int d^2\sigma~\sqrt{h\,} \Big[~
       \left\{~h^{ab} G_{\mu\nu}(X)
             + i\, \epsilon^{ab} B_{\mu\nu}(X)~\right\}
          \partial_a X^\mu \partial_b X^\nu
\nonumber \\
  & \hspace*{8cm} + \alpha'~R\, \Phi(X)~\Big]~.
\label{eq:full-actionE}
\end{align}
We define the partition function for the bosonic string as
\begin{align} 
  & Z[G,\, B,\, \Phi]
  := \int \frac{D_h h\, D_h X}
              { V_{ \text{diffeo.} } \times V_{ \text{Weyl} } }\,
        e^{- S[X,\, h\,;~G,\, B,\, \Phi] }~,
\label{eq:def-partition_function}
\end{align}
where $D_h \bullet$ means that it is invariant measure
with respect to the world sheet metric $h_{ab}$
\footnote{Moreover, $D_h X$ should also be invariant measure
with respect to $G_{\mu\nu}$.}.
By infinitesimal diffeomorphic ($\delta\xi_a$) and
Weyl transformations ($\delta\omega$), the metric changes into
\begin{align}
  & \delta h_{ab}
  = 2\, \delta \omega\, h_{ab} - 2\, \nabla_{(a} \delta\xi_{b)}~,
\label{eq:diffeo-Weyl}
\end{align}
and the volumes of the diffeomorphism and the Weyl transformation
are given by
$V_{ \text{diffeo.} } = \int D_h \xi$ and
$V_{ \text{Weyl} } = \int D_h \omega$, respectively.

According to the standard procedure,
we try to rewrite Eq.(\ref{eq:def-partition_function}) using
\begin{align} 
  & 1 = \Delta_{\text{FP}}[h]~\int D_h \zeta~
              \delta[~h - \hh^\zeta~]~,
\label{eq:def-FP_det}
\end{align}
where $\hh_{ab}$ is a fiducial metric and $\zeta$ means
diffeomorphic ($\xi_a$) and Weyl transformation ($\omega$) parameters,
so that $D_h \zeta = D_h \omega \times D_h \xi$.
From Eqs.(\ref{eq:def-partition_function}) and (\ref{eq:def-FP_det}),
\begin{align} 
   Z[G,\, B,\, \Phi]
  &= \int \frac{ D_h h\, D_h X\, D_h \zeta }
              { V_{ \text{diffeo.} } \times V_{ \text{Weyl} } }~
       \Delta_{\text{FP}}[h]~\delta[~h - \hh^\zeta~]~
        e^{- S[X,\, h\,;~G,\, B,\, \Phi] }
\nonumber \\
  &= \int \frac{ D_{\hh^\zeta} X\, D_{\hh^\zeta} \zeta }
              { V_{ \text{diffeo.} } \times V_{ \text{Weyl} } }~
       \Delta_{\text{FP}}[\hh^\zeta]~
          e^{- S[X,\, \hh^\zeta\,;~G,\, B,\, \Phi] }~,
\label{eq:partition_function}
\end{align}
and from the fact that various quantities are obviously
the world sheet diffeomorphic invariant ones,
the partition function becomes
\begin{align} 
   Z[G,\, B,\, \Phi]
  &= \int \frac{ D_{\hh^\omega} X\, D_{\hh^\omega} \omega
D_{\hh^\omega} \xi }{ V_{ \text{diffeo.} } \times V_{ \text{Weyl} } }~
       \Delta_{\text{FP}}[\hh^\omega]~
          e^{- S[X,\, \hh^\omega\,;~G,\, B,\, \Phi] }
\nonumber \\
  &= \int \frac{ D_{\hh^\omega} X\, D_{\hh^\omega} \omega }
               { V_{ \text{Weyl} } }~
       \Delta_{\text{FP}}[\hh^\omega]~
          e^{- S[X,\, \hh^\omega\,;~G,\, B,\, \Phi] }~,
\label{eq:partition_functionII}
\end{align}
where $\hh^\omega_{ab} := e^{2 \omega}\, \hh_{ab}$ and
we use $\displaystyle{ \int D_{\hh^\omega} \xi = V_{\text{diffeo.}} }$.
Furthermore, Eq.(\ref{eq:partition_functionII}) can be rewritten as
\begin{align} 
   Z[G,\, B,\, \Phi]
  &= \int \frac{ D_{\hh} X\, D_{\hh}\omega }{ V_{\text{Weyl}} }~
       \Delta_{\text{FP}}[\hh]~\times
         \frac{ D_{\hh^\omega} \omega }{ D_{\hh} \omega }~
         \frac{ D_{\hh^\omega} X }{ D_{\hh} X }~
         \frac{ \Delta_{\text{FP}}[\hh^\omega] }
                     { \Delta_{\text{FP}}[\hh] }~
          e^{- S[X,\, \hh^\omega\,;~G,\, B,\, \Phi] }~,
\end{align}
and the Jacobian of the functional volume element produces
the Liouville action
\begin{align} 
  & \frac{ D_{\hh^\omega} \omega }{ D_{\hh} \omega }~
    \frac{ D_{\hh^\omega} X }{ D_{\hh} X }~
    \frac{ \Delta_{\text{FP}}[\hh^\omega] }{ \Delta_{\text{FP}}[\hh] }
  = \exp\left( \frac{D-D_{\text{crit}}+1}{24 \pi}\,
               S_L[\omega,\, \hh] \right)~,
\label{eq:Jacobian} \\
  & S_L[\omega,\, \hh]
  := \int d^2\sigma~\sqrt{\hh}~
   \left( \omega\, \hR - \omega \Hat{\Delta} \omega \right)~,
\label{eq:def-Liouville}
\end{align}
where $D_{\text{crit}} = 26$ is the critical dimension
for the bosonic string and $\Hat{\Delta}$ is
the Laplace-Beltrami operator with respect to $\hh$.
Thus, we obtain
\begin{align} 
  & Z[G,\, B,\, \Phi]
  = \int \frac{ D_{\hh} X\, D_{\hh}\omega }{ V_{\text{Weyl}} }~
       \Delta_{\text{FP}}[\hh]~
           e^{ - I[X,\, \omega,\, \hh\,;~G,\, B,\, \Phi] }~,
\label{eq:partition_functionIII} \\
  & I[X,\, \omega,\, \hh\,;~G,\, B,\, \Phi]
  := S[X,\, \hh^\omega\,;~G,\, B,\, \Phi]
  - \frac{D-D_{\text{crit}}+1}{24\pi}\, S_L[\omega,\, \hh]~.
\label{eq:def-I}
\end{align}

By making use of the equation
$\sqrt{\hh^\omega}\, \hR[\hh^\omega]
= \sqrt{\hh}\, \left( \hR - 2 \Hat{\Delta} \omega \right)$
and gathering $\omega$-dependence, we can rewrite Eq.(\ref{eq:def-I})
as
\begin{align} 
  & I[X,\, \omega,\, \hh\,;~G,\, B,\, \Phi]
  = \calS[X,\, \hh\,;~G,\, B,\, \Phi]
  - \frac{D-D_{\text{crit}}+1}{24\pi}\, S_L[\omega',\, \hh]~,
\label{eq:I-split_form}
\end{align}
where $\omega' := \omega - 6\Phi/(D-D_{\text{crit}}+1)$.
And then, $\calS$ is given by
\begin{align} 
  & \calS
  := \frac{1}{4 \pi \alpha'} \int d^2\sigma~\sqrt{\hh\,}~\left(
       \hh^{ab}\, \tilG_{MN}(X) + \heps^{ab}\, B_{MN}(X) \right)
        \partial_a X^M \partial_b X^N~,
\label{eq:def-calS} \\
  & \tilG_{MN}(X)
  := G_{MN}(X) + \gamma\, \alpha'\,
          \partial_M \Phi(X) \cdot \partial_N \Phi(X)~,
\label{eq:tilG}
\end{align}
where $\gamma := 6/(D - D_{\text{crit}} + 1)$.

Finally, we can equivalently rewrite
Eq.(\ref{eq:partition_functionIII}) in the factored form
\begin{align} 
   Z[G,\, B,\, \Phi]
  &= \int \frac{ D_{\hh}\omega' }{ V_{\text{Weyl}} }~
           e^{ \frac{D-D_{\text{crit}}+1}{24\pi}\,
              S_L[\omega',\, \hh] } \times
    \int D_{\hh} X~\Delta_{\text{FP}}[\hh]~
           e^{ - \calS[X,\, \hh\,;~G,\, B,\, \Phi] }
\nonumber \\
  &= \int \frac{ D_{\hh^\omega} \omega }{ V_{\text{Weyl}} }~
           e^{ \frac{D-D_{\text{crit}}}{24\pi}\,
              S_L[\omega,\, \hh] } \times
    \int D_{\hh} X~\Delta_{\text{FP}}[\hh]~
           e^{ - \calS[X,\, \hh\,;~G,\, B,\, \Phi] }~,
\label{eq:partition_function-splitted}
\end{align}
where we change $\omega'\, \rightarrow\, \omega$ and
the path integral measure,
$D_{\hh}\omega\, \rightarrow\, D_{\hh^\omega}\omega$
in the last equality. 
Thus, the physical modes $X$ decouple from 
the conformal mode of the world sheet metric $\omega$,
so that we can treat the coupling between a test string and the dilaton
in the conformal invariant manner (\ref{eq:def-calS}).

Obviously from the above procedure, the action (\ref{eq:def-calS})
is also applicable for the superstring by replacing 
$D_{\text{crit}} = 26$ with $D_{\text{crit}} = 10$, so that
we have $\gamma = 6/(D-9)$ for the superstring.
\section{NS1-W solution}\label{sec:NS1-W-App}
In this appendix, we summarize the NS1-W solution.
The metric of the $(d+1)$-dimensional target spacetime
for the NS1-W solution
\cite{ref:NS1-W}
is given by
\begin{align} 
  & ds_{d+1}^2 = - \frac{f(r)}{ H_\alpha(r) H_\beta(r) }\, dt^2
  + f^{-1}(r)\, dr^2 + r^2 d\Omega_{d-2}
  + \frac{ H_\beta(r) }{ H_\alpha(r) }\,
      \left( dX^d + \hA_\mu(r)\, dx^\mu \right)^2~,
\label{eq:NS1W-metric-App} \\
  & f(r) = 1 - \left( \frac{r_0}{r} \right)^{d-3}~,
\hspace{1.0cm}
    \hA_\mu (r) dx^\mu
  = - \left( \frac{r_0}{r} \right)^{d-3}\,
    \frac{\sinh\beta\, \cosh\beta}{H_\beta(r)}~dt~,
\label{eq:def-f-hA-App}
\end{align}
where
\begin{align}
  & H_\alpha(r)
  = 1 + \left( \frac{r_0}{r} \right)^{d-3}\, \sinh^2\alpha~,
& & H_\beta(r)
  = 1 + \left( \frac{r_0}{r} \right)^{d-3}\, \sinh^2\beta~.
\label{eq:def-Halpha-Hbeta-App}
\end{align}
The Kalb-Ramond field and the dilaton field are given by
\begin{align}
  & \bmB = \calA_\mu\, dx^\mu \wedge dX^d
  = \left[ - \left( \frac{r_0}{r} \right)^{d-3}
      \frac{ \sinh\alpha\, \cosh\alpha }{ H_\alpha(r) }~dt~\right]
  \wedge dX^d~,
& & e^{-2\Phi} = H_\alpha(r)~,
\label{eq:NS1W-dilaton-App}
\end{align}
and, the charges of $\calA_\mu$ related to the Kalb-Ramond field
and the KK U(1) field $\hA_\mu$ are defined by
\begin{align}
  & Q_\alpha
  := \frac{1}{16 \pi\, G_d} \int_{S^\infty_{d-2}}
     \frac{ e^{-2 \phi} }{k^2}~{}^* \bm{\calF}~,
& & Q_\beta
  := \frac{1}{16 \pi\, G_d} \int_{S^\infty_{d-2}}
     k^2\, e^{-2 \phi}~{}^* \Hat{\bmF}~,
\label{eq:def-hQ}
\end{align}
respectively.
Where $G_d$ is the $d$-dimensional Newton constant related
to the $(d+1)$-dimensional Newton constant $G_{d+1}$ as
$G_d = G_{d+1}/l_z$, and the field $\phi$ is the $d$-dimensional
dilaton field given by $\phi = \Phi - (1/2) \ln k$.
Another important conserved quantity is the ADM mass
and it is easily read off in the Einstein frame $g_{\mu\nu}^{(E)}
:= e^{-4 \phi/(d-2)}\, g_{\mu\nu}$ as
\begin{align}
  & g^{(E)}_{tt}
  \sim -1 + \frac{16 \pi G_d}{(d-2) V_{d-2}}\, \frac{M}{r^{d-3}}
  + \cdots~.
\label{eq:def-Mass}
\end{align}

From Eqs.(\ref{eq:def-hQ}) and (\ref{eq:def-Mass}),
we have for the NS1-W solution
\begin{align}
  & M
  = \frac{l_z\, V_{d-2}}{16 \pi G_{d+1}}\, r_0^{d-3}~\left[~
    d - 2 + ( d - 3 ) \left( \sinh^2\alpha + \sinh^2\beta \right)~
  \right]~,
\label{eq:Mass-ex} \\
  & Q_\alpha
  = ( d - 3 )\, \frac{l_z\, V_{d-2}}{16 \pi G_{d+1}}\,
    r_0^{d-3}~\sinh\alpha\, \cosh\alpha~,
\label{eq:Q_alpha-ex} \\
  & Q_\beta
  = ( d - 3 )\, \frac{l_z\, V_{d-2}}{16 \pi G_{d+1}}\,
    r_0^{d-3}~\sinh\beta\, \cosh\beta~.
\label{eq:Q_beta-ex}
\end{align}
Furthermore, the Hawking temperature and the entropy of the BH
are given by
\begin{align}
  & S_{\text{BH}}
  = \frac{l_z\, V_{d-2}}{4 G_{d+1}}\,
     r_0^{d-2}~\cosh\alpha\, \cosh\beta~,
& & T_{\text{H}}
  = \frac{d-3}{ 4 \pi r_0\, \cosh\alpha\, \cosh\beta }~.
\label{eq:Hawking-T}
\end{align}
%
%

\end{document}